\documentclass[aps,prb,reprint]{revtex4-2}

\pdfoutput=1
\usepackage{amsmath,amssymb,graphicx}
\usepackage{times,txfonts}
\usepackage{hyperref}
\usepackage[utf8]{inputenc}
\usepackage[english]{babel}

\newcommand{\vphast}{{\vphantom *}}
\newcommand{\bk}{\mathbf{k}}
\newcommand{\entrysmall}{${<} \, 10^{-11}$}  

\renewcommand{\Im}{\mathop{\mathrm{Im}}}
\renewcommand{\Re}{\mathop{\mathrm{Re}}}

\begin{document}

\date{August 31, 2022}
\title{Spin oscillations of a single-mode polariton system driven by a plane wave}
\author{S.~S.~Gavrilov}
\affiliation{Institute of Solid State Physics RAS, 142432 Chernogolovka, Russia}
\affiliation{National Research University Higher School of Economics, 101000 Moscow, Russia}

\begin{abstract}
  Theoretical study is performed of a single-mode polariton system
  with linear coupling of spin components.  When combined with an
  ordinary two-particle interaction, the spin coupling involves a
  spontaneous symmetry breaking accompanied by a switch from linear to
  circular polarization under resonant driving.  The asymmetric steady
  states can also lose stability, giving way to oscillatory and
  chaotic dynamics.  Here, we explore a continuous transformation
  between the multistable regime, where the system is steady and
  locked in phase to the pump but has a broken spin symmetry, and
  full-span oscillations of the circular-polarization degree, owing to
  which the symmetry is effectively reestablished.  Such oscillations
  are analogous to the intrinsic Josephson effect and prove to be
  robust against arbitrarily strong perturbations.  Transitional
  phenomena include the Hopf bifurcation, spin bistability of limit
  cycles, and continuous transitions to and from dynamical chaos
  through series of period doubling/halving events.
\end{abstract}

\maketitle

\section{Introduction}

Being coupled exciton-photon pairs, cavity polaritons form coherent
states under resonant excitation~\cite{Kavokin.book.2017, Baas2006,
  Elesin1973, Keldysh2017.en}.  As a rule, the energy level of such a
driven bosonic condensate matches the frequency of the pump wave, in
analogy to a damped oscillator excited by a harmonic force, so that
the condensate is locked in phase to the pump.  Interaction of
polaritons leads to optical multistability which manifests itself in
sharp threshold transitions between steady states in response to
varying excitation parameters~\cite{Baas2004.pra, Gippius2004.epl,
  Carusotto2004, Gavrilov2007.en, Gippius2007, Shelykh2008.prl,
  Gavrilov2010.jetp.en, Paraiso2010, Sarkar2010, Adrados2010,
  Gavrilov2012.bistability, Ballarini2013, Cerna2013}.  However, any
nontrivial dynamics of a simple condensate driven by a plane wave is
naturally hampered.  It is of no surprise that certain ways to release
the phase locking exist for inhomogeneous, relatively complex
polariton systems~\cite{Carusotto2013, Krizhanovskii2010, Pigeon2011,
  Claude2020, Gavrilov2021}.  On the other hand, it was found that
even a single-mode system driven at normal incidence may behave
``freely'', which occurs when all of its locked states get
unstable~\cite{Gavrilov2016, Gavrilov2018, Gavrilov2017.en,
  Gavrilov2020.usp.en}.  Here, we extensively study the onset of such
a global instability, with a focus on a transition from steady states
to self-persistent oscillations depending on the excitation
parameters.

This study is restricted to a system with a few effective degrees of
freedom described by complex-valued amplitudes $\psi_+$ and $\psi_-$
of two spin components.  A similar model was studied in
Ref.~\cite{Sarchi2008} dealing with a double-well Josephson junction
in which two coherent states are coupled by tunneling through a
spatial barrier.  Such a system was shown to have an oscillatory
regime in a particular highly unstable situation when one of two wells
is pumped from the outside but, nevertheless, has a much smaller
steady-state amplitude compared to the other.  Soon afterwards, a
different model with two wells was found to display dynamical
chaos~\cite{Solnyshkov2009.j}; it was relatively complex, however,
being dependent on both the spatial junction and spin coupling and
thus including four rather than two coherent components.  A more
recent example of oscillations in a continuously pumped polariton
system is given by Ref.~\cite{Leblanc2020} which considers a small
micropillar with a nonlinear coupling of two quantized levels.
Finally, a certain combination of resonant and nonresonant excitation
sources can result in a pseudoconservative behavior characterized by a
fully compensated dissipation of polaritons~\cite{Chestnov2021}.

The polariton system considered here is comparatively simple but
exhibits an extremely wide variety of complex dynamical states, both
chaotic and regular.  Theoretically speaking, its key feature is the
spontaneous breakdown of the spin symmetry, which is possible owing to
linear coupling of two spin components~\cite{Gavrilov2020.usp.en}.
Even if they are fully identical and equally pumped, one of two
components is suppressed upon reaching a critical pump amplitude, so
that polaritons acquire very high right- or left-handed circular
polarization~\cite{Gavrilov2013.apl, Sekretenko2013.patterns,
  Gavrilov2014.prb.j}.  In turn, oscillations or dynamical chaos occur
irrespective of the initial conditions only when \emph{all}
steady-state solutions, symmetric and asymmetric, prove to be unstable
for a given pump wave.  This leads us to the problem of finding the
complete set of such solutions.  Earlier, we used to calculate them
numerically for some particular combinations of
parameters~\cite{Gavrilov2016}, which, obviously, does not allow one
to grasp the whole picture.  Here, we report general steady-state
solutions along with explicit formulae for the spectra of elementary
excitations.

In contrast to Refs.~\cite{Gavrilov2016} and \cite{Gavrilov2018} which
demonstrated sharp switches from steady states to chaos, this work is
devoted to a continuous transition between different, mostly regular,
dynamical regimes in the course of varying pump amplitude~$f$.
Namely, we have found that steady condensate states with broken
symmetry may continuously turn into limit cycles.  Since there are two
spin states feasible under the same external conditions, the
corresponding limit cycles are also twinned in phase space.  They are
completely distinct near the bifurcation point; however, expansion of
a given periodic orbit upon varying $f$ makes it approach the twin
one, which opens the possibility of switches between two spin domains
and leads to their unification.  After a pair of alternative orbits
have merged, they run through the whole range of circular-polarization
degrees from $-1$ to $+1$ and represent a new, purely dynamical
symmetric state which is stable against arbitrary perturbations.

In what follows, the base model is introduced in Sec.~\ref{sec:model}.
Section~\ref{sec:ms} is aimed at finding the complete set of
fixed-point solutions, after which we determine the conditions of
their bifurcations into nonsteady states.  Section~\ref{sec:lc} deals
with a variety of periodic and chaotic dynamical states and their
continuous transformations.  Section~\ref{sec:discussion} contains a
general discussion and concluding remarks.

\section{Base model}
\label{sec:model}

The coherently excited system of cavity polaritons obeys the following
wave equations~\cite{Kavokin.book.2017},
\begin{equation}
  \label{eq:gp}
  i \hbar \frac{\partial \psi_\pm}{\partial t} =
  \left( E_0 - i \gamma
    + V \psi_\pm^* \psi_\pm^\vphast
  \right) \psi_\pm^\vphast
  + \frac{g}{2} \psi_\mp^\vphast
  + f_\pm^\vphast e^{-i E_p t / \hbar},
\end{equation}
where $\psi_+$ and $\psi_-$ are complex-valued amplitudes of two spin
components, $E_0$ and $\gamma$ are the polariton energy and decay
rate, $g$ is the spin coupling rate, $V$ is the polariton-polariton
interaction constant, $f_+$ and $f_-$ are the pump amplitudes
corresponding to the right- and left-handed polarizations of light,
$E_p / \hbar$ is the pump frequency that is assumed to be close to the
polariton resonance.  The pump wave vector is zero, which corresponds
to normal incidence of the driving wave.

For a planar two-dimensional system, $\psi_\pm$ depend on space and
time, whereas $E_0 = E_0(-i \hbar \nabla)$ is an operator determined
by the dispersion law of the lower polariton branch.  For a
zero-dimensional system representing a small micropillar, $E_0$ means
the ground level of size quantization.  On the assumption that all
other levels play no role, the equations are projected onto the
corresponding spatial eigenstate; as a result, $\psi_+$ and $\psi_-$
depend only on time.  The same equations describe an effectively
spinless system with two quantum wells~\cite{Sarchi2008}.  As shown in
Ref.~\cite{Leblanc2020}, similar equations also hold for a spinless
system in a micropillar when one takes into account the interaction of
two quantized levels; however, their coupling rate $g$ appears to be
amplitude-dependent.

In case if $f_+ = f_- = 0$ and $\gamma = 0$, model~(\ref{eq:gp}) is
reduced to a simple Josephson junction.  The eigenlevels are split
($E = E_0 \pm g/2$ for $|\psi_\pm| \to 0$), therefore the phase
difference of two components varies with time and involves
oscillations.  Among other bosonic systems (\cite{Josephson1962,
  Cataliotti2001, Levi2007}), such phenomenon is displayed by a freely
evolving pair of coupled polariton condensates~\cite{Lagoudakis2010,
  Abbarchi2013, Ohadi2016}.  The spin oscillations in a pointlike
system are referred to as the \emph{intrinsic} Josephson
effect~\cite{Shelykh2008.prb.j, Shelykh2010, Gavrilov2014.prb.j} owing
to exactly the same form of the equations.

Everything is changed if the system is driven directly, i.\,e., $f_+$
or $f_-$ is nonzero.  In this case Eqs.~(\ref{eq:gp}) have solutions
of the form $\psi_\pm(t) = \bar\psi_\pm e^{-i E_p \hbar / t}$.  The
Josephson oscillations are no longer possible since both components
oscillate at the same ``forced'' frequency.  Time-independent
amplitudes $\bar\psi_+$ and $\bar\psi_-$ obey a set of algebraic
equations
\begin{align}
  \label{eq:fp-plus}
  \left( D + i \gamma - V |\bar\psi_+|^2 \right) \bar\psi_+
  - \frac{g}{2} \bar\psi_- &= f_+,\\
  \label{eq:fp-minus}
  \left( D + i \gamma - V |\bar\psi_-|^2 \right) \bar\psi_-
  - \frac{g}{2} \bar\psi_+ &= f_-,
\end{align}
where $D = E_p - E_0$ (pump detuning).  Such solutions are usually
called \emph{steady states}, or \emph{fixed points}, because $E_p$ can
be taken as zero without loss of generality.

The dissipative character of the system $(\gamma > 0)$ is crucially
important.  Owing to dissipation, all solutions would be guaranteed to
have the forced frequency~$E_p / \hbar$ at $t \to \infty$ if the
interaction constant $V$ were zero.  At the same time, nonlinearity
does not automatically lead to a nonsteady behavior.  If $V > 0$ but
$g = 0$, then even a two-dimensional model~(\ref{eq:gp}) with many
degrees of freedom evolves to a plane-wave state oscillating at
$E = E_p$; this final state is asymptotically stable, so that all weak
perturbations decay with time.

Thus, the fixed-point states are the ``most common'' attractors of the
phase trajectory for $f_\pm \neq 0$ and
$\gamma > 0$~\cite{Gavrilov2020.usp.en}.  A
different---nonsteady---type of solutions comes into being only when
all fixed points become unstable.

\section{Fixed-point states and their bifurcations}
\label{sec:ms}

Let us solve Eqs.~(\ref{eq:fp-plus}), (\ref{eq:fp-minus}) at
$f_+ = f_- = f$.  Notice that in this case the model is
spin-symmetric.  Since the right-hand sides of the two equations are
the same, one can equate the left-hand sides and have
\begin{equation}
  \label{eq:ratio}
  \frac{\bar\psi_-}{\bar\psi_+} =
  \frac
  {a + i \gamma - \mu_+}
  {a + i \gamma - \mu_-},
\end{equation}
where $\bar\psi_+ \neq 0$ and
\begin{equation}
  \label{eq:designations_1}
  \mu_\pm = V |\bar\psi_\pm|^2, \quad a = D + \frac{g}{2}.
\end{equation}
Taking the absolute square of~(\ref{eq:ratio}) yields
\begin{equation}
  \label{eq:mu12}
  \mu_+^3 - \mu_-^3 - 2a (\mu_+^2 - \mu_-^2)
  + (a^2 + \gamma^2)(\mu_+ - \mu_-) = 0.
\end{equation}
When considering the asymmetric states with $\mu_+ \neq \mu_-$, one
may cancel $\mu_+ - \mu_-$ from~(\ref{eq:mu12}), which results in
\begin{equation}
  \label{eq:asymm_gen}
  w^2 - (u - a)^2 = \gamma^2,
\end{equation}
where
\begin{equation}
  \label{eq:designations_2}
  u = \mu_+ + \mu_-, \quad w = \sqrt{\mu_+ \mu_-}.
\end{equation}
Equation~(\ref{eq:asymm_gen}) suggests that one of two spin components
vanishes at $u = a$ and $\gamma \to 0$.  In general, not all
combinations of $u$ and $w$ are possible in accordance with their
definition.  Indeed,
\begin{equation}
  \label{eq:limitation}
  0 < (\mu_+ - \mu_-)^2 = u^2 - 4w^2 =
  u^2 - 4 \, \bigl[ (u - a)^2 + \gamma^2 \bigr].
\end{equation}
The inequality is satisfied for $l_1 < u < l_2$, where
\begin{equation}
  \label{eq:limits}
  l_{1,2} = \frac23 \Bigl( 2a \mp \sqrt{a^2 - 3\gamma^2} \Bigr).
\end{equation}
The ``most asymmetric'' state with $u \approx a$ is seen to be inside
the allowed interval, given that $D$ and $g$ are positive and
sufficiently great compared to $\gamma$.  This condition is always
assumed hereafter.

In order to find $u$ as a function of $f,$ one can start with finding
the reverse dependence $f = f(u)$.  To do so, we multiply both sides
of Eqs.~(\ref{eq:fp-plus}) and (\ref{eq:fp-minus}) by their respective
complex conjugates and then sum up the two equations.  The result
depends on $\mu_\pm$ and
$\bar\psi_+^* \bar\psi_-^\vphast + \bar\psi_-^* \bar\psi_+^\vphast$;
however, the latter quantity also can be expressed in terms of
$\mu_\pm$ using~(\ref{eq:ratio}).  Specifically,
\begin{multline}
  \label{eq:real12}
  \bar\psi_+^* \bar\psi_-^\vphast + \bar\psi_-^* \bar\psi_+^\vphast =
  \left[ (a - \mu_+)(a - \mu_-) + \gamma^2 \right]
  \\ {} \times
  \left[
    \frac{\mu_+}{(a - \mu_-)^2 + \gamma^2} +
    \frac{\mu_-}{(a - \mu_+)^2 + \gamma^2}
  \right].
\end{multline}
The symmetry between $\mu_+$ and $\mu_-$ suggests that all $\mu$ can
be rewritten in terms of $u$ and $w^2$, after which $w^2$ is easily
eliminated using~(\ref{eq:asymm_gen}).  In so doing, after a lengthy
calculation one arrives at the final expression:
\begin{equation}
  \label{eq:res}
  |f|^2 =  \frac{1}{V} \left(
    (2a - u) \, (D - u)^2 + \frac{\gamma^2 \, (2D - u)^2}{2a - u}
  \right)
\end{equation}
for $u \in (l_1, l_2)$.  The singularity point $u = 2a$ is never
reached, because $l_2 < 2a$ so long as $\gamma > 0$.
Eq.~(\ref{eq:res}) can be reduced to a quartic equation for $u$ which
is solvable in a general way.  However, the main features of the
solutions can be understood using the already obtained reverse
dependence~$f(u)$.  Below, without loss of generality we assume $f$ to
be real-valued and positive.

Consider a region around $u \sim a$.  If $\gamma$ is noticeably
smaller than $a$, one can disregard the second term in~(\ref{eq:res}).
Therefore, equation $df / du = 0$, which determines the turning points
of the solutions, becomes quadratic.  It has the following roots,
\begin{equation}
  \label{eq:turning-roots}
  u_1 = 2a - (D + g) = D, \quad u_2 = 2a - \frac{D + g}{3},
\end{equation}
whereas the corresponding values of $f^2$ are
\begin{equation}
  \label{eq:turning-f}
  f_1^2 = 0, \quad f_2^2 = \frac{4}{27 \, V} (D + g)^3.
\end{equation}
The interval from $\max \{u_1, l_1\}$ to $u_2$, wherein function
$u(f)$ is single-valued and has positive derivative, is the main
domain of asymmetric solutions which will be further analyzed in this
work.

\begin{figure}[!b]
  \centering
  \includegraphics[width=\linewidth]{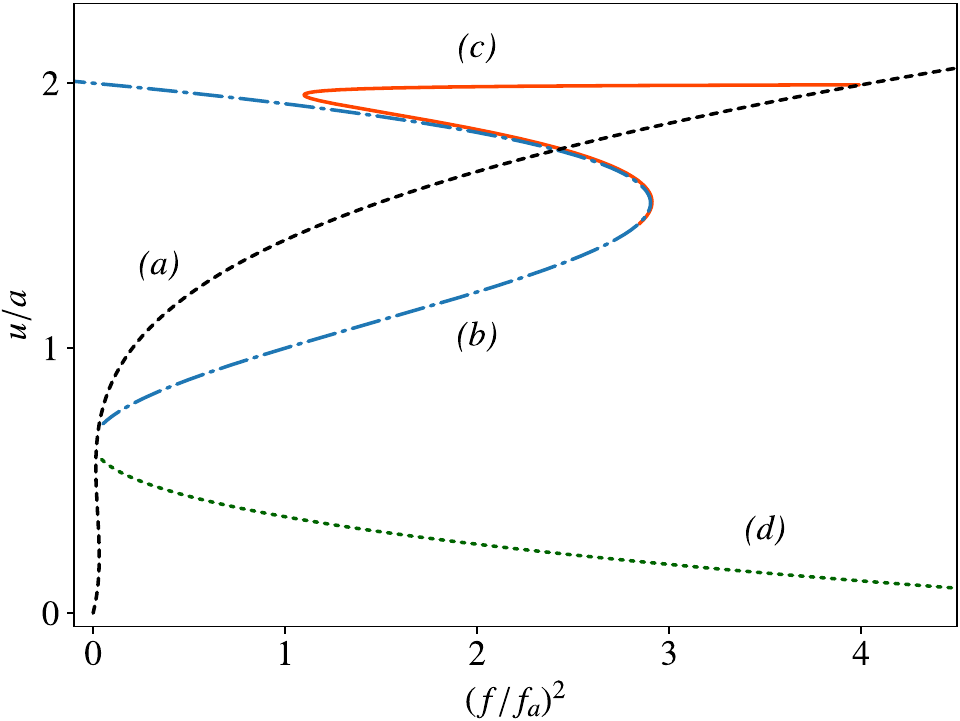}
  \caption{\label{fig:1} Fixed-point states in case of a
    spin-symmetric excitation.  (a)~symmetric states~(\ref{eq:symm})
    with $\bar\psi_+ = \bar\psi_-$; (b)~approximate asymmetric
    solution $f^2(u) = (2a - u)(D - u)^2 / V$; (c)~exact asymmetric
    solution~(\ref{eq:res}) in a region where it is noticeably
    different from (b); (d)~spurious branch $(u < l_1)$.  Parameters:
    $g = 9 \gamma$, $D = 8.1 \gamma$; $f_a = ag^2/4V$.}
\end{figure}

Figure~\ref{fig:1} exemplifies the full set of the fixed-point
solutions obtained at $g = 9 \gamma$ and $D = 0.9 g$.  First, it
contains the spin-symmetric branch with $\bar\psi_+ = \bar\psi_-$ for
each~$f$ (curve \textit{a}).  The corresponding expression for
$f^2(u)$ is easily obtained by taking the absolute square of
Eq.~(\ref{eq:fp-plus}) or (\ref{eq:fp-minus}), which yields
\begin{equation}
  \label{eq:symm}
  f^2_\mathrm{symm} = \frac{u}{2V} \biggl[
    \left( D - \frac{g}{2} - \frac{u}{2} \right)^2 + \gamma^2
  \biggr].
\end{equation}
This formula reproduces the well-known S-shaped curve of a spinless
polariton system~\cite{Baas2004.pra} with $V|\psi|^2 = u/2$ and pump
detuning $D - g/2$ (the latter quantity equals the difference between
$E_p$ and the eigenlevel whose polarization matches that of the pump
wave).

Also shown in Fig.~\ref{fig:1} are the approximate asymmetric solution
obtained from (\ref{eq:res}) by dropping the second term
(curve~\textit{b}), the exact solution in the vicinity of $u = 2a$
(curve~\textit{c}), and the spurious branch for $u < l_1$
(curve~\textit{d}).  Branches \textit{b} and \textit{d} are seen to
combine into an upside-down S-shaped curve.  Indeed, if we substitute
$u = 2a - u'$ in (\ref{eq:res}), then for $\gamma \to 0$ we once again
arrive at the usual response function describing a spinless system
whose effective $V |\psi|^2$ and pump detuning are equal to $u'$ and
$D' = D + g$, respectively, which is also evident from
Eqs.~(\ref{eq:turning-roots}) and (\ref{eq:turning-f}).  Notice that
$D'$ is quite large and exceeds even the pump detuning from the
farther eigenlevel (i.\,e., $D + g/2$).  Since
$f_2^2 - f_1^2 \propto D'^3$, function $u(f)$ changes very slowly
throughout the main asymmetric domain $(l_1, u_2)$.

\begin{figure}[!b]
  \centering
  \includegraphics[width=\linewidth]{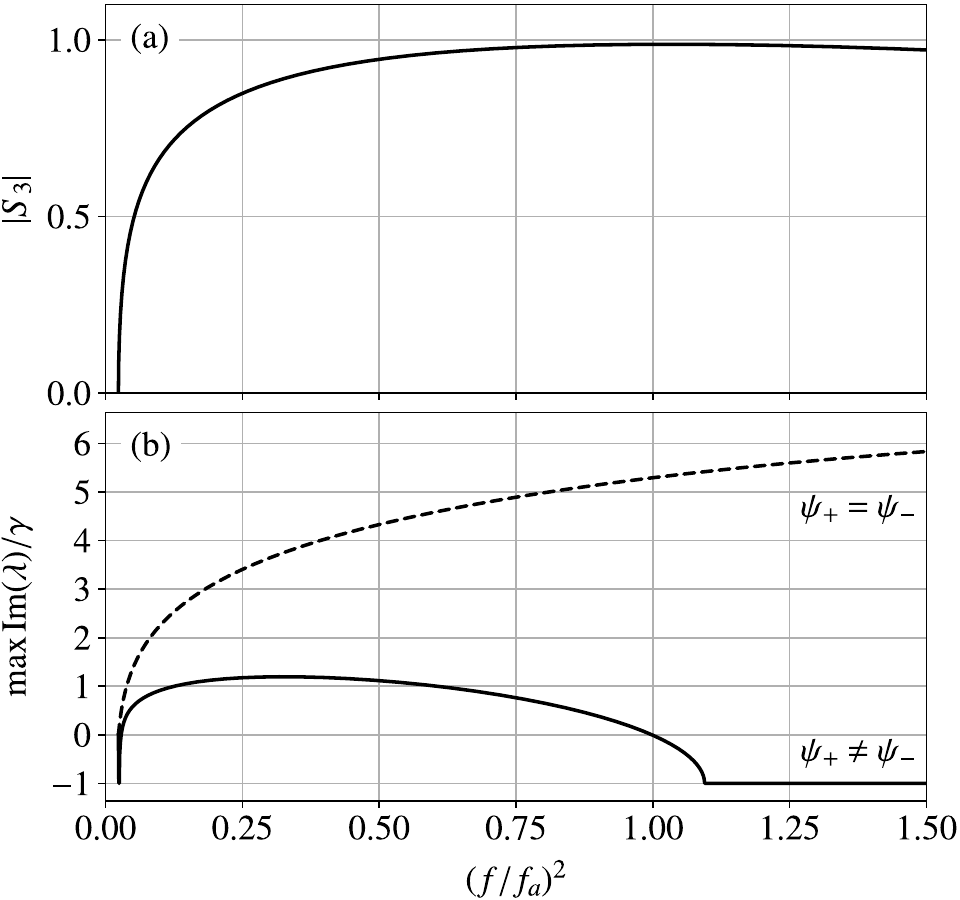}
  \caption{\label{fig:2} (a) Circular-polarization degree as a
    function of pump intensity.  (b) Corresponding decay rates in the
    symmetric (dashed line) and main asymmetric (solid line) states of
    the system.  $\lambda$ denotes the eigenvalues
    of~(\ref{eq:thematrix}).  Parameters are the same as in
    Fig.~\ref{fig:1}.}
\end{figure}

Let us find the circular-polarization degree, also referred to as the
third Stokes parameter $(S_3)$.  Proceeding from its definition and
then using (\ref{eq:designations_2}) and (\ref{eq:asymm_gen}), we have
\begin{equation}
  \label{eq:s3}
  S_3
  = \frac{\mu_+ - \mu_-}{\mu_+ + \mu_-}
  = \pm \sqrt{1 -
    \frac{4 \, \bigl[ (u - a)^2 + \gamma^2 \bigr]}
    {u^2}}
\end{equation}
in the allowed range of asymmetric states $l_1 < u < l_2$.  Notice
that $S_3(l_{1,2})= 0$ [see~(\ref{eq:limitation})], so that both
limiting points belong to the symmetric branch~(\ref{eq:symm}).  The
maximum of $|S_3|$ equals $1 - 2 \gamma^2 / a^2$, given that
$\gamma \ll a$, and is reached at $u = a + \gamma^2 / a$.
Figure~\ref{fig:2}(a) represents the dependence of $|S_3|$ on $f^2$
for $u$ in a subinterval of $(l_1, u_2)$.  The system is seen to have
nearly total right- or left-handed polarization in a wide range of
$f^2$ around $f_a^2 = ag^2 / 4V$ (this quantity equals~(\ref{eq:res})
at $u = a$ and $\gamma \to 0$).

In a similar manner one deals with an arbitrary function of
$\bar\psi_\pm$.  One calculates $S_3$ and then
$\mu_\pm = (u/2) (1 \pm S_3)$ depending on $u$, after which the
relative phase of $\bar\psi_+$ and $\bar\psi_-$ is determined using
(\ref{eq:ratio}) or (\ref{eq:real12}).  Any function of $\bar\psi_\pm$
is thereby expressed in terms of $u$ and also related to $f^2$,
because $u$ is a single-valued function of $f^2$ in the considered
domain.  Here we employ this procedure to find the spectrum of
elementary excitations (also known as Bogolyubov
quasiparticles~\cite{Bogolyubov1947.en,Pitaevskii.book}) for a
fixed-point state with given $\bar\psi_\pm$.  The eigenenergies and
eigenstates of the excitations are determined by the following
$4 \times 4$ matrix,
\begin{widetext}
  \begin{equation}
    \label{eq:thematrix}
    \mathcal L(\bar\psi_\pm) =
    \begin{pmatrix}
         E_p - i\gamma + 2V |\bar\psi_+|^2 - D_\bk
       & V \bar\psi_+^2
       & g/2
       & 0
      \\ -V \bar\psi_+^{*2}
       & E_p - i\gamma - 2V |\bar\psi_+|^2 + D_\bk
       & 0
       & -g/2
      \\ g/2
       & 0
       & E_p - i\gamma + 2V |\bar\psi_-|^2 - D_\bk
       & V \bar\psi_-^2
      \\ 0
       & -g/2
       & -V \bar\psi_-^{*2}
       & E_p - i\gamma - 2V |\bar\psi_-|^2 + D_\bk
    \end{pmatrix},
  \end{equation}
\end{widetext}
where $D_\bk$ means $E_p - E_0(\bk)$ for a spatially extended system
characterized by the dispersion law $E_0(\bk)$ or merely equals $D$ in
the single-mode case.  The derivation of~(\ref{eq:thematrix}) from
Eq.~(\ref{eq:gp}) can be found elsewhere~\cite{Sarchi2008,
  Gavrilov2017.en, Gavrilov2020.usp.en}.  Solving the characteristic
equation $\det (\mathcal L (\bar\psi_\pm) - \lambda \mathrm I) = 0$
yields four eigenvalues~$\lambda$.  The considered fixed-point state
is said to be \emph{unstable} when at least one of them has positive
imaginary part for any~$\bk$, which implies spontaneous growth of
infinitesimal fluctuations in the corresponding $k$-state or in the
driven mode itself.  If, by contrast, all $\Im \lambda$ are negative,
fluctuations decay with time and the solution is stable.

Figure~\ref{fig:2}(b) represents the dependence of
$\Gamma = \max \Im \lambda$ on $f^2$ in a subinterval of $(l_1, u_2)$.
It is seen that all symmetric solutions (\ref{eq:symm}) are unstable.
In fact, they lose stability already at $u = l_1$, where the
asymmetric branch originates, and remain unstable until it eventually
merges back, which occurs at $u = l_2$ and $f^2 = 4f_a^2$.  Thus, when
the asymmetric solutions merely exist in accordance with
(\ref{eq:limitation}), the symmetry is \emph{really} destroyed in the
sense of spontaneously growing fluctuations.  Such an ``exchange of
stabilities'' between two branches of fixed points is indicative of
the \emph{transcritical bifurcation}~\cite{Strogatz.book.2015}.  This
kind of transition takes place at $g \ge D$; in the opposite case, the
$u = l_1$ point is not directly reachable and the transition is
discontinuous.

The spin symmetry breakdown under coherent driving was experimentally
observed in a microcavity with $g / \gamma \sim 2$, where it resulted
in a fast transition from linear to circular polarization of the
emitted light~\cite{Gavrilov2013.apl, Sekretenko2013.patterns,
  Gavrilov2014.prb.j}.  After the symmetry has broken, such a system
gets locked in a state with very high $|S_3|$ similar to
Fig.~\ref{fig:2}(a), so that a significant increase in pump intensity
is required for the transition to the upper symmetric state.  Later it
was found that the asymmetric solutions may also lose stability,
provided that $g / \gamma \gtrsim 4$~\cite{Gavrilov2016,Gavrilov2018}.
The corresponding Bogolyubov spectra were found analytically in the
case when one of two spin components nearly vanishes owing to symmetry
breaking~\cite{Gavrilov2017.en}.  In Fig.~\ref{fig:2}(b) we draw
$\Gamma$ in a wide area of $f$ and now intend to offer a simple
qualitative explanation of the presented result.  (The explicit
formulae for all eigenvalues of~(\ref{eq:thematrix}) are given in
\hyperref[sec:appendix]{Appendix}.)

\begin{figure}[!b]
  \centering
  \includegraphics[width=\linewidth]{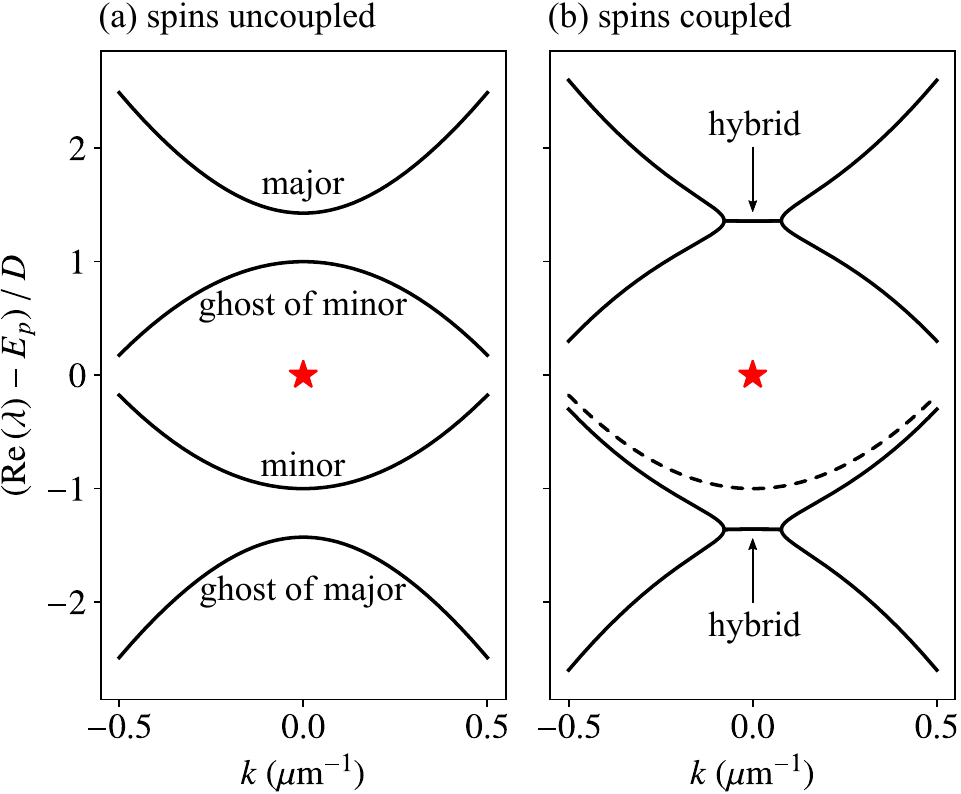}
  \caption{\label{fig:3} Hybridization of elementary excitations.
    $\lambda$ denotes the eigenvalues of~(\ref{eq:thematrix}).
    Parameters in (b) correspond to the asymmetric solution in
    Fig.~\ref{fig:2} at $f = f_a$.  In (a), all parameters are the
    same except $g = 0$.  The dispersion law corresponds to a
    GaAs-based microcavity.}
\end{figure}

First, let us take Eq.~(\ref{eq:gp}) with $g = 0$, so that spins are
uncoupled.  The excitations are then determined by two $2 \times 2$
submatrices of~(\ref{eq:thematrix}).  Solving the characteristic
equations gives
\begin{equation}
  \label{eq:exc_spinless}
  \lambda_{1,2} = E_p - i \gamma \mp \sqrt{(D_\bk - 2\mu)^2 - \mu^2}
\end{equation}
for each spin component.  For the purposes of illustration, consider a
spatially extended system with a nearly parabolic dispersion law
$E_0(k)$ in the vicinity of $k = 0$.  If $\mu \to 0$, then
$\lambda_1(k)$ equals $E_0(k) - i\gamma$, whereas $\lambda_2(k)$
represents the ``ghost'' branch that originates owing to the pairwise
interaction~\cite{Kohnle2011,Stepanov2019}.  As $\mu$ is increased,
the normal dispersion branch is blueshifted and the ghost branch with
$d^2 E / dk^2 < 0$ is redshifted evenly.  Now suppose that one of two
spin components is strong and the other is zero; then the normal
branch of the major component can be close to resonance with the ghost
branch of the minor component and vice versa, which is sketched in
Fig.~\ref{fig:3}(a).  If $g = 0$, this is nothing but an approximate
coincidence of energies; however, a nonzero spin coupling acts to
hybridize both resonant pairs below and above the pumped mode as shown
in Fig.~\ref{fig:3}(b).  Owing to hybridization, the joint states
acquire positive $\Gamma$ and grow
spontaneously~\cite{Gavrilov2020.usp.en}.

The above consideration enables one to estimate the optimal parameters
where the instability of the asymmetric states is the most strong.
The normal level of the major component is determined by
(\ref{eq:exc_spinless}) and equals
$E_p + \sqrt{(D - 2\mu)^2 - \mu^2}$, whereas the ghost level of the
minor component is $E_p + D$ (for $k = 0$).  The two levels coincide
when $3\mu = 4D$; then a formal replacement of $\mu$ with
$a = D + g/2$ yields $D = 3g/2$.  Such a relation between $D$ and $g$
leads to nearly optimal hybridization conditions for the ``most
asymmetric'' state with $u \approx a$ even when the spin coupling is
introduced fully self-consistently.

Let us turn back to Fig.~\ref{fig:2}(b) showing the dependence of
$\Gamma$ on $f^2$ in the asymmetric domain.  Since $D = 0.9 g$, the
hybridization conditions are not optimal for $u = a$; namely, the
uncoupled major level seen in Fig.~\ref{fig:3}(a) is noticeably higher
than the nearest ghost level.  In the discussed example, $\Gamma$
turns to zero at $f \approx 0.999 f_a$, so that further increasing $f$
cancels the hybridization completely, and, vice versa, decreasing $f$
lowers the major level and thus strengthens its coupling with the
ghost one.  That is why $\Gamma$ shows an increase with decreasing
$f^2$ from $f_a^2$ down to ${\sim} \, 0.3 f_a^2$.  In the range of
small $f^2$ around $0.1 f_a^2$, the ghost level is also lowered, which
is explained by the population of the minor spin component in view of
partially restored symmetry.  Since both the major and ghost levels
decrease simultaneously, they remain coupled and ensure positive
$\Gamma$.  As a result, the asymmetric states are unstable until $f^2$
becomes as small as ${\sim} \, 0.03 f_a^2$.

In summary, we have ascertained that \emph{all} fixed-point solutions
can be unstable in a broad range of pump intensities.  The asymmetric
solutions change very sharply near the lesser of two $\Gamma = 0$
points.  By contrast, the greater one, found in the vicinity of
$f = f_a$ for $D \sim g$, appears to be a ``smooth'' bifurcation point
in which the steady states continuously arise from or turn into the
oscillatory states.  Such a transition, referred to as the \emph{Hopf
  bifurcation,} is analyzed below.

\section{Limit cycles}
\label{sec:lc}

\begin{figure}
  \centering
  \includegraphics[width=\linewidth]{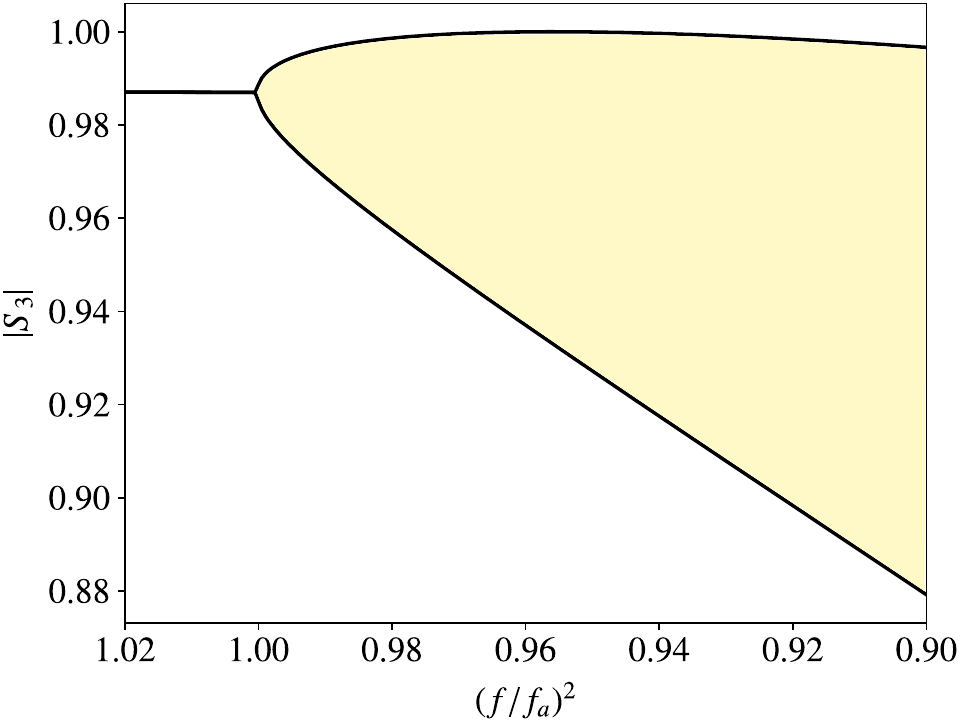}
  \caption{\label{fig:4} The Hopf bifurcation in the vicinity of
    $f = f_a$.  The two lines represent the minimum and maximum of
    $|S_3|$ of an established solution depending on~$f$.  The initial
    conditions play no role.}
\end{figure}

Figure~\ref{fig:4} displays a series of numeric solutions of
Eqs.~(\ref{eq:gp}) in the vicinity of the Hopf bifurcation.  The
system parameters correspond to Figs.~\ref{fig:1} and \ref{fig:2}.
Since the upper $\Gamma = 0$ point of the asymmetric branch is very
close to $f_a = \sqrt{a g^2 / 4V}$, we keep using the latter quantity
as a natural unit of~$f$ (in general, the $\Gamma = 0$ points are
sensitive to $D/g$ and not pinned to $f_a$).  The series shown in
Fig.~\ref{fig:4} consists of many independent solutions obtained for
different $(f / f_a)^2$ around 1 with a step of $5 \times 10^{-4}$.
Each solution was recorded after a 100~ns long ``establishment''
period that largely exceeds the characteristic time of transient
processes $\hbar / \gamma$.  The results do not depend on the initial
conditions, sharp jumps accompanying symmetry breaking and everything
else occurred at the early stages of the evolution.

The continuous onset of the oscillations in the $\Gamma = 0$ point is
not self-evident.  In spite of the fact that crossing such points is
indeed very likely to result in the second-order phase transitions in
lasers~\cite{Haken1975,Haken.book.1983,Cross1993}, the polariton
system driven above resonance can behave differently.  For instance,
the parametric scattering, which starts smoothly while the pumped mode
resides at the lower branch of the S-shaped curve, is accompanied by a
gradual accumulation of energy under constant external conditions
until the system eventually jumps to the upper
branch~\cite{Gavrilov2014.prb.b,Gavrilov2015,Gavrilov2021}.  The
considered system, by contrast, has no alternative fixed points it
could tend towards, whereas the filling of the hybrid modes does not
act to increase the total energy.  (Indeed, one of two spin components
is suppressed and the other has already reached its higher-energy
state, so that an additional blueshift would only worsen resonance
conditions for the pumped mode.)  Thus, the parametric scattering is
inherently balanced for each $f$ in the vicinity of $f = f_a$.

\begin{figure}
  \centering
  \includegraphics[width=\linewidth]{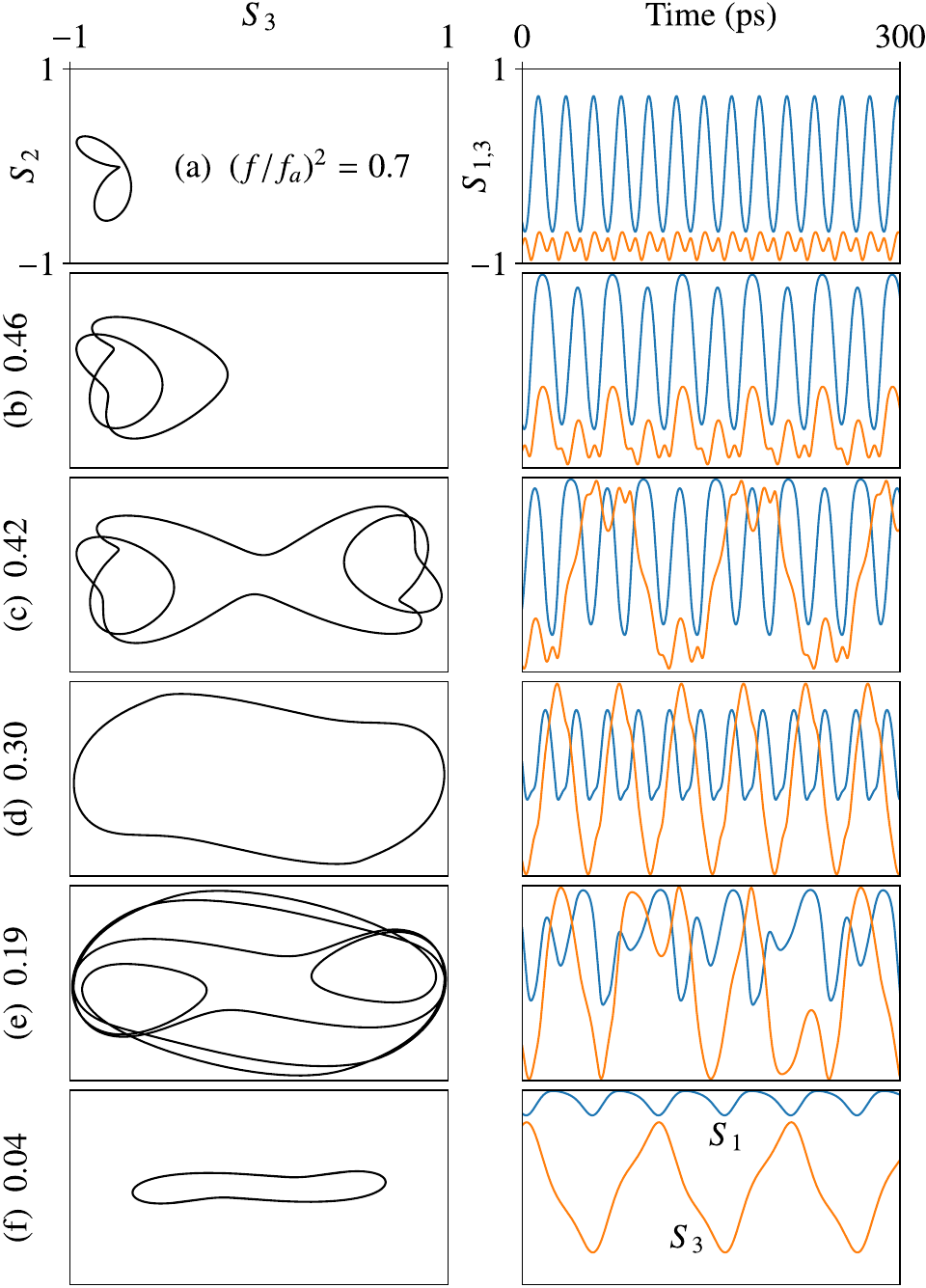}
  \caption{\label{fig:5} Limit-cycle solutions at different
    $(f / f_a)^2$.  Quantities $S_{1,2,3}$ are the Stokes vector
    components.  The solutions are bistable for (a) and (b) and
    independent of the initial conditions for (c--f), so that the sole
    basin of attraction covers the entire phase space.}
\end{figure}

Let us now decrease $f$ and enter the range of more intensive
oscillations. Figure~\ref{fig:5} displays the trajectories (left side)
and explicit dynamics (right side) in terms of the Stokes vector
components $S_{1,2,3}$.  Along with the circular-polarization degree
$S_3$ [(\ref{eq:s3})], they include
\begin{equation}
  \label{eq:s12}
  S_1 = \frac{|\psi_x|^2 - |\psi_y|^2}{|\psi_x|^2 + |\psi_y|^2},
  \quad S_2 = \frac{\psi_x^*\psi_y^{\vphantom *}
    + \psi_y^*\psi_x^{\vphantom *}}
  {|\psi_x|^2 + |\psi_y|^2}.
\end{equation}
$S_{1,2}$ represent the degrees of linear polarization in two spatial
bases, $(\mathbf{e}_x, \mathbf{e}_y)$ and
$\frac{1}{\sqrt{2}} (\mathbf{e}_x + \mathbf{e}_y, \mathbf{e}_x -
\mathbf{e}_y)$.  In accordance with the usual definition,
\begin{equation}
  \label{eq:xy}
  \dbinom{\psi_x}{\psi_y} = \frac{1}{\sqrt{2}}
  \left(
    \begin{matrix}
      1 & 1 \\
      i & -i
    \end{matrix}
  \right)
  \dbinom{\psi_+}{\psi_-},
\end{equation}
so that all spin-symmetric states with $\psi_+ = \psi_-$ must be
polarized along the $x$ axis and have $S_1 = +1$.  The Stokes vector
$\mathbf{S}$ defined this way has the length of unity.

Figure~\ref{fig:5}(a) exemplifies the periodic solutions obtained in a
broad range of $(f/f_a)^2$ from 0.9 to 0.5.  Apart from growing
amplitude, they are qualitatively the same for each~$f$.  The system
is seen to oscillate near the state with circular polarization, either
right- or left-handed.  Such solutions prove to be stable against weak
perturbations and represent two alternative periodic attractors (limit
cycles).  In a sense, they continue the two respective branches of
fixed points.

The amplitude of oscillation grows with decreasing~$f$.  At
$(f/f_a)^2 = 0.46$ [Fig.~\ref{fig:5}(b)], the orbit nearly approaches
the midpoint of spin domains and becomes more complex; being composed
of two different circuits, it shows a doubled period.  The next step
in decreasing $f$ makes the system enter the opposite domain, which
means the possibility of spin switches.  Such switches can be chaotic
in a certain transitional range of $f$, but eventually the alternative
solutions unite and the common orbit runs through both spin domains
evenly [Fig.~\ref{fig:5}(c)].  As $f$ is decreased further, the
solution undergoes a sequence of transformations
[Figs.~\ref{fig:5}(d)--(f)], ending up in a fixed-point state with
$S_1 = +1$.

\begin{table}
  \begin{tabular}{|c||c|c|c|c|c|c|c|}
    \hline
    $n$ & $-3$ & $-2$ & $-1$ & 0 & $+1$ & $+2$ & $+3$
    \\ \hline
    $\mu_x / a$
    & \entrysmall
    & 0.06
    & \entrysmall
    & 0.52
    & \entrysmall
    & 0.18
    & \entrysmall
    \\ \hline
    $\mu_y / a$
    & 0.03
    & \entrysmall
    & 0.35
    & \entrysmall
    & 0.11
    & \entrysmall
    & 0.02
    \\ \hline
  \end{tabular}
  \caption{\label{tab:1} Discrete spectrum of a periodic solution at
    $(f/f_a)^2 = 0.3$ [Fig.~\ref{fig:5}(d)].  The energy levels are
    located at $E = E_p + n \Delta$, where
    $\Delta \approx 1.036 \, D/2$ and $n$ is integer.  Quantities
    $\mu_{x,y}$ are equal to $V |\psi_{x,y}|^2$.}
\end{table}

As soon as the opposite solutions unite, the mean values of $|\psi_+|$
and $|\psi_-|$ become the same.  Together with strict periodicity,
this gives rise to discrete energy spectra whose individual levels
have linear polarizations.  Table~\ref{tab:1} represents such a
spectrum for $(f/f_a)^2 = 0.3$.  The energy dependences were obtained
by the Fourier transform of $\psi_{x,y}(t)$ over an interval of 40~ns.
All spectral lines, which are fairly unbroadened in view of their
``parametric'' nature, exhibit alternating linear polarizations
oriented along the $x$ and $y$ axes.  The behavior of the solution
resembles the intrinsic Josephson effect in a system with two
orthogonally polarized levels \cite{Shelykh2008.prb.j, Shelykh2010,
  Gavrilov2014.prb.j}, which manifests itself in varying phase
difference $\arg (\psi_x^* \psi_y^\vphast)$ and, thus, in the
oscillations between $S_2 = \pm 1$ and $S_3 = \pm 1$, similarly to
Fig.~\ref{fig:5}(d).  However, in our case the alternating states are
numerous.  The higher-order levels get populated through a direct
``off-branch'' scattering~\cite{Savvidis2001,Whittaker2005} and decay
exponentially with increasing $|n|$.  By contrast, the peaks listed in
Table~\ref{tab:1} are the main ones which reflect an inherent balance
of instabilities.  The difference between adjacent co-polarized
levels, $2\Delta$, corresponds to the oscillation period
$T = \pi \hbar / \Delta$.  Since the total intensity is a very slow
function of $f$ in the asymmetric domain, $2\Delta$ is still
comparable to $D$ in accordance with Fig.~\ref{fig:3}, despite the
changed polarization states.  The solutions shown in
Figs.~\ref{fig:5}(c), (e), and (f) also have unbroadened spectra with
alternating $x$ and $y$ levels, but their $\Delta$'s are markedly
different because of the period-doubling effect.

All orbits shown in Figs.~\ref{fig:5}(c)--(f) appear to be the only
attractors existing for the respective values of~$f$.  Each orbit is
stable against arbitrary perturbations and does not depend on the
initial conditions, which was checked for a broad range of $\psi_+$
and $\psi_-$.  Namely, the initial values of $|\psi_\pm|$ were
independently varied in the interval from 0 to $1.5 \sqrt{a}$ with a
step of $0.1 \sqrt{a}$, whereas $\arg (\psi_+^* \psi_-^\vphast)$ was
varied from 0 to $2 \pi$ with a step of $0.1 \pi$.  When the accuracy
was sufficiently high, all trajectories for each given $f$ merged into
the same orbit.

\begin{figure}[!b]
  \centering
  \includegraphics[width=\linewidth]{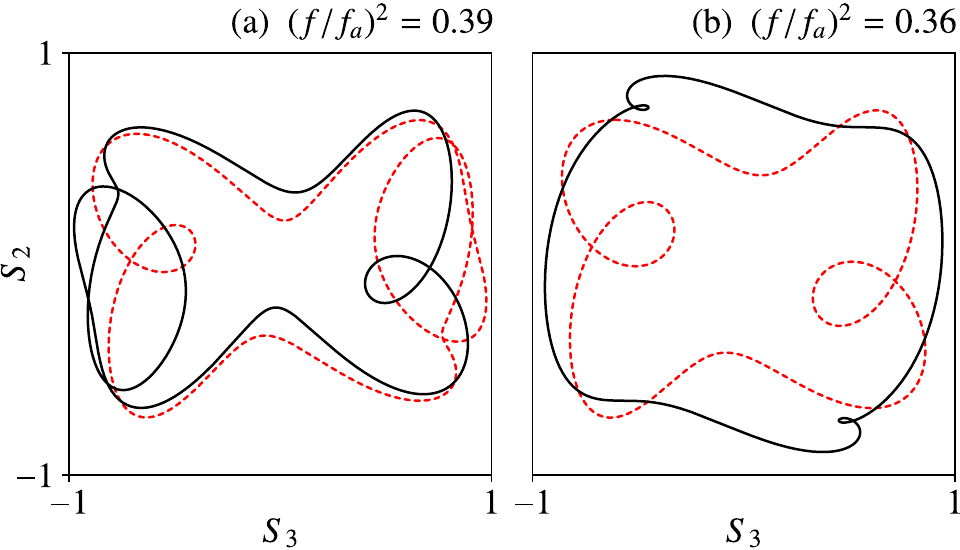}
  \caption{\label{fig:6} Bistable periodic solutions.  The solid and
    dashed curves (also distinct in color) display alternative
    trajectories which are chosen depending on the initial
    conditions.}
\end{figure}

Being qualitatively different, the absolutely stable orbits are
separated by more complex dynamical regimes in the transitional ranges
of~$f$ (Fig.~\ref{fig:6}).  First, the system can lose the spin
symmetry even after the unification of spin domains.  As a result, a
pair of solutions with broken symmetry become equally possible
depending on the initial conditions.  One such pair, obtained at
$(f/f_a)^2 = 0.39$, is shown in Fig.~\ref{fig:6}(a).  A less common
situation with two spin-symmetric solutions is found at
$(f/f_a)^2 = 0.36$ [Fig.~\ref{fig:6}(b)].  In both cases the system
can be switched between its alternative periodic states using
additional pulsed excitation.

Apart from bistable limit cycles, the system displays dynamical chaos.
In particular, chaotic solutions are found at $(f/f_a)^2 \sim 0.44$,
close to the fusion of spin domains, where trajectories are extremely
sensitive to fluctuations.  The way from Fig.~\ref{fig:5}(d) to
\ref{fig:5}(e) also contains a chaotic interval.  Specifically, the
system loses the spin symmetry in analogy to Fig.~\ref{fig:6}(a) and
then experiences a continuous transition to chaos through a number of
the period-doubling events.  The initial oscillation period $T_0$ is
comparable to $2 \pi \hbar / D$, similarly to Fig.~\ref{fig:5}(d).  As
$f$ is decreased, the period is doubled at certain critical points
$f_{i}$, some of which are listed in Table~\ref{tab:2}.

\begin{table}[!h]
  \begin{tabular}{|c||c|c|c|c|}
    \hline
    $i$ & 1 & 2 & 3 & 4
    \\ \hline
    $T_i / T_0$ & 2 & 4 & 8 & 16
    \\ \hline
    $f_i^2 / f_a^2$ & 0.2509 & 0.2426 & 0.2410 & 0.24068
    \\ \hline
    $\displaystyle \frac{f_{i-1}^2 - f_{i-2}^2}{f_i^2 - f_{i-1}^2}$
    & --- & --- & 5.19 & 5.00
    \\ \hline
  \end{tabular}
  \caption{\label{tab:2} Several events of period doubling.}
\end{table}

At each step, the interval between successive bifurcations becomes
$\delta_i \approx 5$ times shorter than previously.  This behavior is
general for one-dimensional discrete chaotic systems such as the
logistic map~\cite{Feigenbaum1978}.  With increasing $i$, the
shortening coefficients $\delta_i$ tend to the universal limit
$\delta = 4.669\!\ldots,$ known as the Feigenbaum constant, that does
not depend on a specific system or the choice of its bifurcation
parameter.  The continuous systems with at least three degrees of
freedom demonstrate a wider variety of routes to chaos, but the
period-doubling scenario remains one of the most
common~\cite{Bohr1998}.  It should only be noticed that finding the
bifurcation points with high $i$ can be difficult, because it requires
a rapidly growing computational accuracy.

\begin{figure}
  \centering
  \includegraphics[width=\linewidth]{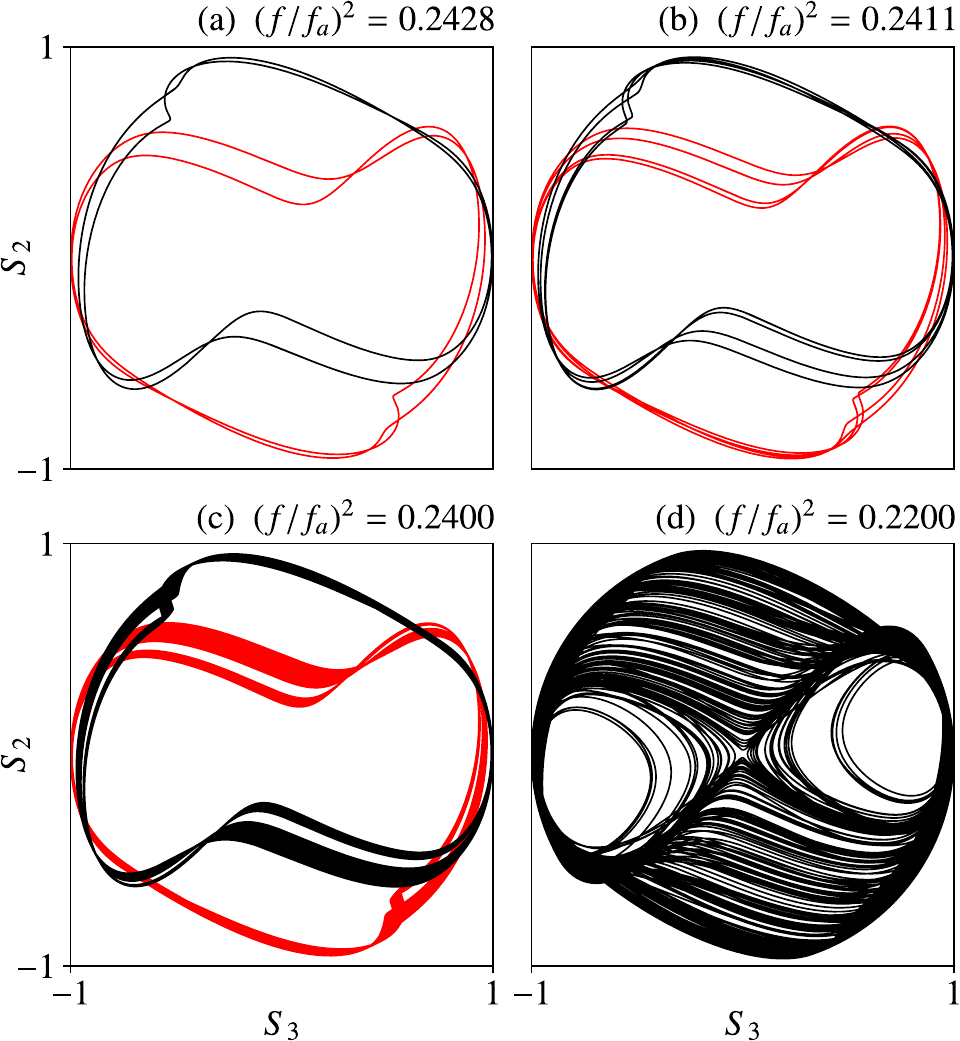}
  \caption{\label{fig:7} Transition to chaos through a cascade of the
    period-doubling bifurcations.  Different colors in (a)--(c)
    represent alternative solutions.  The time span in (c) and (d) is
    10~ns.}
\end{figure}

Figures~\ref{fig:7}(a,b) display the solutions obtained after the
first and second period-doubling events.  The orbits have the
correspondingly increased numbers of circuits but remain qualitatively
similar.  Since the interval between successive bifurcations is small,
each newborn pair of circuits cannot diverge very strongly by the
moment of the next doubling.  The solutions retain self-similarity in
spite of many bifurcations even when they become chaotic at
$f \lesssim f_{i \, {=} \, \infty}$ [Fig.~\ref{fig:7}(c)].  As a
result, such solutions remain bistable.  The alternative chaotic
states seen in Fig.~\ref{fig:7}(c) originate from the same initial
conditions as a pair of their respective ancestors at the top of the
bifurcation cascade.  However, further decreasing $f$ makes the
trajectories unite and thus cancels bistability [Fig.~\ref{fig:7}(d)].
Afterwards, the \emph{reverse} bifurcation cascade occurs, which
comprises successive period halvings, ending up in a state shown in
Fig.~\ref{fig:5}(e).  In spite of a complicated orbit, the latter
state is particularly stable in a broad range of $(f/f_a)^2$ from 0.19
to 0.09.  In other words, a ``multiple'' periodicity of the trajectory
is no longer related to a reduced stability.

Solving Eqs.~(\ref{eq:gp}) in a general form is hardly possible.
Nevertheless we can explain very roughly, which kind of solutions
should be expected depending on $g$ and $D$.  The main empirical
parameter is the maximum gain rate of the asymmetric fixed points,
$\max_f \Gamma$.  In our examples, it is comparable
to~$\gamma$~[Fig.~\ref{fig:2}(b)].  Notice that both $u$ and $\Gamma$
are slow functions of $f$ for the asymmetric branches.  At the same
time, $\max_f\Gamma$ strongly depends on $D/g$ and reaches its own
maximum at $D/g \sim 3/2$ (see Sec.~\ref{sec:ms}), where it appears to
be significantly greater than $\gamma$.  In this case the interval of
$f$ which contains no stable fixed points is mostly ``chaotic''.  By
contrast, at $D/g \lesssim 0.5$ all nonsteady solutions, if they ever
exist, are regular and do not even exhibit bistability.  The other
``regular'' area is found at $D/g \gtrsim 2$.  Here, $\Gamma$ is
decreased for the usual reason: the hybridization conditions in the
sense of Fig.~\ref{fig:3} are hampered by a comparatively great energy
mismatch of the excitations, but now the upper major level turns out
to be too low.  This area of $D/g$ has its own interesting features,
particularly for the case of a spatially extended system when the
hybridization conditions are better satisfied for nonzero wave
numbers.  At last, the area of $D \sim g$, on which we have been
focused in the present work, exhibits several regular and several
chaotic intervals of $f$ and thus delivers an especially rich
phenomenology.

We have not yet discussed the role played by the ratio $g/\gamma$.  It
was chosen to be equal to 9 in all examples, but in our general
analysis we only supposed it to be indefinitely great.  On one hand,
this assumption is well justified, because the instability of the
asymmetric fixed points is known to be possible for all
$g/\gamma \gtrsim 4$ (the greater, the better)
\cite{Gavrilov2017.en,Gavrilov2020.usp.en}.  The symmetric states lose
stability in the interval $l_1 < u < l_2$ that is also independent of
$\gamma$ when $\gamma \ll a$ [see (\ref{eq:limits})].  At the same
time, there exists a different kind of asymmetric solutions,
represented by curve $c$ in Fig.~\ref{fig:1}.  It is sensitive to
$\gamma$ and completely absent at $\gamma = 0$ when the second term in
(\ref{eq:res}) equals zero.  Owing to this term, equation $df/du = 0$
has an additional root $f=f_3$ that determines the beginning point of
the respective branch.  If $\gamma$ is small compared to $g$ and $D$,
then~\cite{Gavrilov2020}\footnote{The derivation of $f_3$ in
  Ref.~\cite{Gavrilov2020} was based on a different approximation that
  led to $f_3^2 \approx 2 \gamma a g / V$.  In turn,
  formula~(\ref{eq:f3}) is based on Eq.~(\ref{eq:res}) and represents
  an exact solution of the equation $df / du = 0$ at $\gamma \to 0$.}
\begin{equation}
  \label{eq:f3}
  f_3^2 \approx \frac{2 \gamma g (g + D)}{V}.
\end{equation}
All new solutions with $du/df > 0$ are stable; thus, being feasible at
a given $f$, they break up the ``exclusive'' character of limit
cycles.  The considered case of $g/\gamma = 9$ is in fact close to a
state in which $f_3 \propto \sqrt{\gamma}$ becomes less than $f$ at
the upper $\Gamma = 0$ point of the asymmetric domain for $D \sim g$.
When this indeed occurs for a somewhat greater $g/\gamma$, the Hopf
bifurcation turns into a more complex discontinuous transition at
$f = f_3$, which proceeds between the new fixed points and nonzero
limit cycles.  The new solutions bring about interesting phenomena
such as phase bistability; namely, a pair of alternative steady states
may have equal polarizations $(\mathbf{S})$ but opposite phases and
thus annihilate each other at the places of spatial contact.  This
leads to a perfectly spontaneous formation of topological excitations
(quantized vortices and dark solitons) in spatially extended polariton
systems~\cite{Gavrilov2020}.

\section{Discussion}
\label{sec:discussion}

Among all known polariton systems which show nontrivial dynamics under
one-mode driving, the discussed system appears to be the most simple,
at least from a conceptual viewpoint.  It does not require even the
spatial extent and does not rely upon any sort of nonlinear
interaction between two coherent components.  The very possibility of
chaos in this system was doubted in earlier
works~\cite{Solnyshkov2009.j}.  Nevertheless, an unexpectedly rich set
of nonsteady states, both regular and chaotic, was found.

The reason for the dynamical complexity of the considered system lies
in the fact of its spontaneously broken symmetry.  The asymmetric
fixed points easily lose stability, giving way to nontrivial dynamics.
Therefore, the challenge is to find out where exactly the solutions of
Eqs.~(\ref{eq:fp-plus})(\ref{eq:fp-minus}) become unstable.  In
general, the total number of different fixed points that coexist for
the same $f$ can reach \emph{nine}, which makes the whole picture
quite complicated.  That is why our preceding
study~\cite{Gavrilov2016} did not go far beyond a mere demonstration
of oscillatory and chaotic regimes for some particular parameters.  In
this work, we have found that the intrinsic symmetry of
Eqs.~(\ref{eq:fp-plus})(\ref{eq:fp-minus}) enables one to solve them
exactly at $f_+ = f_-$ and thus find all critical points for the
transitions between the multistable and essentially nonsteady states.
The calculation of the $\Gamma = 0$ points is performed
straightforwardly, based on the energies of elementary excitations
derived in \hyperref[sec:appendix]{Appendix}.

As a result, we have investigated nonsteady polariton states in a wide
range of system parameters and found several different types of spin
oscillations.  The first occurs when a pair of asymmetric fixed points
turn into infinitesimal limit cycles which thus continuously extend
the respective branches of the multistability diagram.  In analogy to
conventional multistability, the system can be switched between two
opposite trajectories by means of a pulsed perturbation of the driving
field.  Another kind of oscillations comes into being when the spin
symmetry is \emph{reestablished} via unification of opposite limit
cycles.  The corresponding energy spectra consist of a number of
equally spaced unbroadened lines with alternating orthogonal
polarizations.  These solutions closely resemble the Josephson effect
in freely evolving Bose systems.  Notice, however, that the main pair
of spectral lines corresponds to the Bogolyubov modes rather than
polariton eigenlevels, which means that the discussed effect has no
linear analogues.

The periodic variation of $S_3$ (average spin) covers almost the whole
range from $-1$ to $+1$.  Such states with a ``dynamical'' spin
symmetry can be completely independent of the initial conditions and
robust against any temporal perturbations.  At the same time, they
remain very sensitive to the continuous-wave excitation parameters
and, for instance, exhibit discrete switches, doublings or halvings,
of the oscillation period upon varying $f$.  Being highly controllable
already on the scale of $10^2$~ps (in GaAs-based samples), this type
of coherent states can be used for information encoding and
transmission.  Finally, we have found that in a certain range of pump
amplitudes the system experiences a deterministic and continuous
transition to chaos through an infinite cascade of period doublings.
The chaotic variation of $S_3$ represents an extremely fast analogue
of the \emph{polarization chaos} formerly observed in electrically
driven laser diodes~\cite{Virte2013.natphoton, Virte2013.pra,
  Sciamanna2015}.

For simplicity, our study has been restricted to the case of
zero-dimensional polariton systems in micropillars.  At the same time,
the analysis of the fixed points and their bifurcations remains valid
in a more general case of laterally uniform cavities.  Owing to the
spatial extent, the simultaneous instability of all plane-wave
solutions for a given $f$ can result in spontaneous ordering,
spatiotemporal oscillations or chaos, and \emph{chimera states} in
which ordered and disordered subsystems coexist~\cite{Gavrilov2016,
  Gavrilov2018, Gavrilov2020.usp.en, Gavrilov2020}.  A precise spin
symmetry and homogeneity of the model enable one to solve the
steady-state equations analytically but do not constitute rigid
requirements to the experimental conditions.  However, a significant
strength of the spin coupling compared to the decay rate
$(g / \gamma > 4)$ remains necessary, still being a challenging task
for experimental realization.

\acknowledgements

I am grateful to V.~D.~Kulakovskii and N.~N.~Ipatov for fruitful
discussions.  The work was supported by the Russian Science
Foundation, grant No.\ 19-72-30003.

\appendix*

\section{Explicit formulae for energies of elementary excitations}
\label{sec:appendix}

Here we calculate the eigenvalues of~(\ref{eq:thematrix}) for all
steady states obeying Eqs.~(\ref{eq:fp-plus}), (\ref{eq:fp-minus}).
Owing to the intrinsic symmetry of the equations, one can reduce this
multiparameter problem to a single free variable. The results apply to
both a pointlike micropillar and a homogeneous cavity in which a
plane-wave polariton condensate is excited at $\bk = 0$.  In the
latter case, $D_\bk$ means $E_p - E_0(\bk)$, where $E_0(\bk)$ is the
usual dispersion law of polaritons.

The characteristic equation for~(\ref{eq:thematrix}),
$\det(\mathcal L(\bar\psi_\pm) - \lambda \mathrm{I}) = 0$, has the
following roots,
\begin{equation}
  \label{eq:app:E:gen}
  \lambda = E_p - i \gamma \pm \sqrt{\frac{P \pm \sqrt{Q}}{2}},
\end{equation}
where
\begin{align}
  P &= 3 (\mu_+^2 + \mu_-^2) - 4D_\bk (\mu_+ + \mu_-)
    + 2D_\bk^2 + \frac{g^2}{2},
  \label{eq:app:P:gen}
  \\ \notag Q &= g^2 \left\{
  4 \left[ D_\bk - (\mu_+ + \mu_-) \right]^2
  -(\mu_+ + \mu_-)^2 + 4 \mu_+ \mu_- \cos^2 \varphi
  \right\}
  \\ &+ (\mu_+ - \mu_-)^2
  \left[ 4 D_\bk - 3(\mu_+ + \mu_-) \right]^2,
  \label{eq:app:Q:gen}
\end{align}
$\mu_\pm = V|\bar\psi_\pm|^2$, and
$\varphi = \arg(\bar\psi_+^*\bar\psi_-^{\vphantom +})$.  To obtain
$\cos \varphi$ for $\bar\psi_+ \neq \bar\psi_-$, recall that
(\ref{eq:real12}) can be unambiguously expressed in terms of
$u = \mu_+ + \mu_-$ and $w = \sqrt{\mu_+ \mu_-}$; this leads to
\begin{equation}
  \label{eq:app:cos}
  w \cos \varphi = a - u + \frac{2\gamma^2}{2a - u}.
\end{equation}
(In case of multimode systems, quantity $a$ corresponds to the driven
mode, i.\,e., $a \equiv D_{\bk \, {=} \, 0} + g/2$.)  We assume that
$l_1 < u < l_2$ and employ Eq.~(\ref{eq:asymm_gen}), which enables us
to rewrite all functions of $\mu_+$, $\mu_-$, and $\varphi$ in terms
of only one ``condensate'' variable,~$u$. As a result,
(\ref{eq:app:E:gen}) takes the following general form for the states
with spontaneously broken symmetry,
\begin{multline}
  \label{eq:app:E:asymm}
  \lambda^\mathrm{(a)} =
  E_p - i \gamma \pm \frac{1}{\sqrt{2}} \Biggl\{
  2(D_\bk - u)^2 + u^2 - 6(a - u)^2 - 6\gamma^2 +
  \frac{g^2}{2}
  \\ {} \pm \Biggl[
  (u^2 - 4(a - u)^2 - 4\gamma^2)
  (4 D_\bk - 3u)^2
  + 4 g^2 (D_\bk - u)^2 - g^2 u^2
  \\ {} + 4 g^2
  \left( a - u + \frac{2 \gamma^2}{2a - u} \right)^2
  \Biggr]^{1/2} \Biggr\}^{1/2}.
\end{multline}
A characteristic example for $\Re \lambda^\mathrm{(a)}(k)$ is shown in
Fig.~\ref{fig:3}(b).  An approximate particular case of
(\ref{eq:app:E:asymm}), obtained for the state in which one of two
spin components is exactly zero, was considered earlier in
Refs.~\cite{Gavrilov2017.en,Gavrilov2018}.

In order to find the excitations for the spin-symmetric solutions, one
should set $\varphi = 0$ and $\mu_+ = \mu_- = u/2$ in
(\ref{eq:app:P:gen}), (\ref{eq:app:Q:gen}), which yields
\begin{equation}
  \label{eq:app:E:symm}
  \lambda^\mathrm{(s)} =
  E_p - i \gamma \pm \frac{1}{2} \sqrt{[2(D_\bk - u) \pm g]^2 - u^2}.
\end{equation}
As expected, expressions (\ref{eq:app:E:asymm}) and
(\ref{eq:app:E:symm}) become exactly the same in both limiting points
$u = l_{1,2}$.

The two signs of $g$ in the radicand of Eq.~(\ref{eq:app:E:symm})
correspond to a pair of orthogonally polarized states (e.\,g., the
plus sign leads to an effective $D' = D + g/2$ which is the pump
detuning from the lower resonance level having the $y$ polarization,
thus, the respective Bogolyubov modes are also $y$-polarized).  When
$u = l_1$, the degenerate solutions $\lambda^{(\mathrm{s}, y)}$ and
$\lambda^\mathrm{(a)}$ have exactly zero
$\Gamma \equiv \max \Im \lambda$.  This quantity becomes negative at
$u > l_1$ for $\lambda^\mathrm{(a)}$ and positive for
$\lambda^{(\mathrm{s}, y)}$, which implies spontaneous growth of the
$y$-polarized Bogolyubov excitations on the background of the
$x$-polarized condensate and necessarily leads to the spin symmetry
breaking.  If $g > D$, then the $u = l_1$ point lies on a stable
branch with a positive slope $(du/df > 0)$, so that the polarization
transition at $f = f(l_1)$ occurs continuously.

\end{document}